\documentclass[conference]{IEEEtran}
\IEEEoverridecommandlockouts
 
\usepackage{graphicx}
\usepackage{amsmath}
\usepackage{booktabs}
\usepackage{url}
\usepackage{xcolor}
\usepackage{enumitem}
\usepackage{listings}
\usepackage{algorithm}
\usepackage{algpseudocode}
\usepackage{multirow}
\usepackage{amssymb}
\usepackage{makecell}
\usepackage{tikz}

\usetikzlibrary{fit,positioning,arrows.meta}
\usepackage[most,listings]{tcolorbox}
\usepackage[font=small,labelfont=bf]{caption}

\definecolor{MotivBg}{RGB}{247,248,250}
\definecolor{MotivFrame}{RGB}{180,180,180}
\definecolor{LineNumGray}{RGB}{140,140,140}
\definecolor{SnagRed}{RGB}{170,50,50}
\definecolor{bggray}{RGB}{245,245,245}
\definecolor{framegray}{RGB}{180,180,180}

\newcommand{\refine}[1]{{\color{black}#1}}

\newcommand{\approach}{\textsc{Eager}}

\setlist[itemize]{leftmargin=*, topsep=1pt, itemsep=1pt}
\setlist[enumerate]{leftmargin=*, topsep=1pt, itemsep=1pt}
\renewcommand{\arraystretch}{1.0}
\lstdefinestyle{motivation}{
  language=Python,
  basicstyle=\ttfamily\footnotesize,
  numbers=left,
  numberstyle=\tiny\color{LineNumGray},
  numbersep=6pt,
  frame=none,
  tabsize=2,
  breaklines=true,
  showstringspaces=false,
  commentstyle=\itshape\color{gray!70}
}

\newtcblisting{motivationbox}{
  listing only,
  listing options={style=motivation},
  colback=MotivBg,
  colframe=MotivFrame,
  arc=3pt,
  boxrule=0.6pt,
  left=12pt,
  right=6pt,
  top=4pt,
  bottom=4pt,
}

\def\BibTeX{{\rm B\kern-.05em{\sc i\kern-.025em b}\kern-.08em
    T\kern-.1667em\lower.7ex\hbox{E}\kern-.125emX}}
\begin{document}
\title{Executing as You Generate: Hiding Execution Latency in LLM's Code Interpreters}

\author{
\IEEEauthorblockN{Zhensu Sun\IEEEauthorrefmark{1}}
\IEEEauthorblockA{Singapore Management University\\
Singapore\\
zssun@smu.edu.sg}
\and
\IEEEauthorblockN{Zhihao Lin\IEEEauthorrefmark{1}}
\IEEEauthorblockA{Beihang University\\
China\\
mathieulin@buaa.edu.cn}
\and
\IEEEauthorblockN{Zhi Chen}
\IEEEauthorblockA{Singapore Management University\\
Singapore\\
zhi.chen.2023@smu.edu.sg}
\and
\IEEEauthorblockN{Chengran Yang}
\IEEEauthorblockA{Singapore Management University\\
Singapore\\
cryang@smu.edu.sg}
\and
\IEEEauthorblockN{Mingyi Zhou}
\IEEEauthorblockA{Beihang University\\
China\\
zhoumingyi@buaa.edu.cn}
\and
\IEEEauthorblockN{Li Li}
\IEEEauthorblockA{Beihang University\\
China\\
lilicoding@ieee.org}
\and
\IEEEauthorblockN{David Lo}
\IEEEauthorblockA{Singapore Management University\\
Singapore\\
davidlo@smu.edu.sg}
\thanks{\IEEEauthorrefmark{1}These authors contributed equally to this research.}
}
\maketitle

\begin{abstract}
Current LLM systems are increasingly equipped with a code interpreter that executes generated code to obtain results.
This works serially: the model first generates the complete code, then an interpreter executes it.
This sequential workflow leaves the executor idle during generation and the generator idle during execution, resulting in unnecessary end-to-end latency.
Our key observation is that an LLM, unlike a human developer, emits code tokens left to right and does not backtrack over what it has already written.
This makes it possible to start executing a piece of code while later tokens are still being generated.
We formalize this \textbf{parallel execution} paradigm, modeling it as a three-stage pipeline of generation, detection, and execution, and derive closed-form latency bounds that characterize its speedup potential and operating regimes.
We then present \approach{}, a concrete implementation featuring AST-based chunking, dynamic batching with gated execution, and early error interruption.
We evaluate \approach{} across four benchmarks, seven LLMs, and three execution environments.
The overlap mechanism hides almost all execution behind generation, reducing the non-overlapped portion of execution time by up to 99.8\% and cutting end-to-end latency by up to 37.3\% on error-free runs.
\end{abstract}

\section{Introduction}
\label{sec:intro}

Modern LLM systems increasingly act not by producing text alone but by
producing code that is run on the user's behalf.
A growing class of systems hands the model a code interpreter: the model writes a self-contained program, an interpreter executes it, and the result is returned to the user or fed back to the model for the next step.
This pattern underlies commercial tools such as Anthropic's Claude code execution tool~\cite{anthropic_code_execution} and OpenAI's Code Interpreter~\cite{openai_code_interpreter}, as well as research systems such as CodeAct~\cite{wang2024codeact} and Open Interpreter~\cite{openinterpreter2024}, where executable Python program is the unit of action.
In these settings the program is usually a short, single-file script that runs a piece of data analysis or produces a plot, and the user waits for the whole writing-and-running process to finish before seeing a result.

This workflow is serial, where the model first generates the complete program, and only after the final token is emitted does execution begin.
While the model is generating, the interpreter sits idle.
While the interpreter runs, the model has nothing left to do.
The user-perceived latency is therefore the sum of two phases that run back to back, even though running an early Python statement never requires the later code to be generated first.

\begin{figure}[t]
\centering
\includegraphics[width=\linewidth]{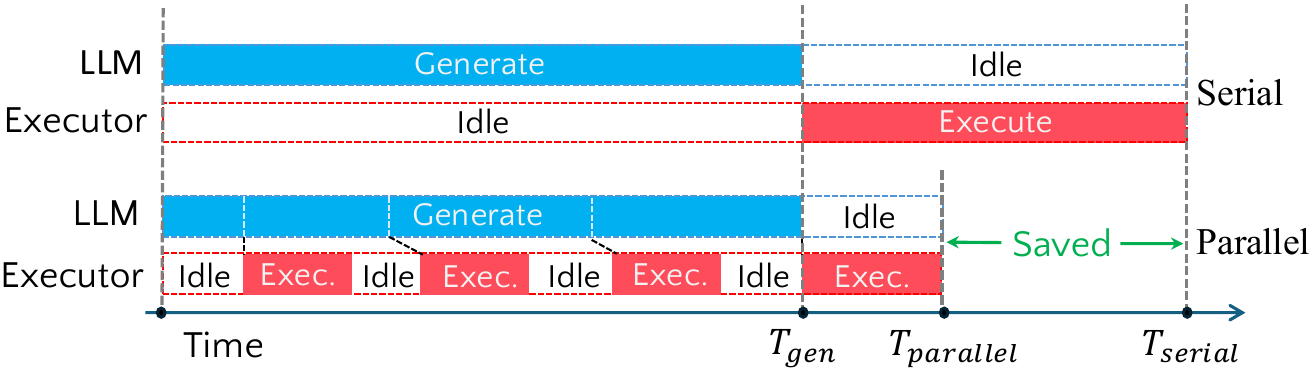}
\caption{An illustrative example comparing Serial Execution and Parallel Execution. For a code snippet with four chunks, Parallel Execution overlaps the first three chunks with the generation process, saving the corresponding waiting time.}
\label{fig:overview}
\end{figure}

Our starting observation is a property of how these models write code.
A human developer revises constantly, jumping back to edit earlier lines, so a program is only safe to run once the developer decides it is done.
A token-based language model does not work this way.
Under standard autoregressive decoding, it emits tokens from left to right and never rewrites a token once it has been produced, so every prefix of the output is already final.
As soon as the model has emitted a complete statement, that statement is exactly what will appear in the finished program, and there is no
reason to wait for the rest of the code before running it.
We call this idea \emph{parallel execution}: each statement is dispatched to the interpreter the moment it is complete, so that execution of earlier statements overlaps with generation of later ones.
Figure~\ref{fig:overview} illustrates the
effect.
Serial execution costs roughly $T_{gen} + T_{exec}$, whereas parallel
execution costs roughly $T_{gen} + T_{tail}$, where $T_{\text{gen}}$, $T_{\text{exec}}$, and $T_{\text{tail}}$ denote the generation time, the total execution time, and the execution time of only the final chunk, respectively.

To the best of our knowledge, the parallel paradigm between the execution and generation of LLM-produced code has not yet been explored in the research community.
Prior work has proposed incremental execution strategies~\cite{openinterpreter2024,yang2024intercode}, where a model generates a few lines, executes them, and conditions subsequent generation on the observed output.
For example, Open Interpreter~\cite{openinterpreter2024} executes code in Jupyter-style cells and feeds the output back to the model, and EG-CFG~\cite{lavon2025execution} integrates real-time execution signals into the code generation process.
These approaches use execution feedback to improve code quality by steering subsequent generation.
However, they are still fundamentally serial with respect to latency: the model must pause generation while awaiting each execution result, and the total wall-clock time includes both generation and execution in full.
In contrast, the parallel execution paradigm we propose targets a complementary goal: reducing user-perceived latency by overlapping generation and execution, without altering the generated code.
As a result, the following question remains open: Is parallel execution of LLM-generated code practically viable, and what are its benefits and costs?

To realize this paradigm, we present \approach{} (\textbf{E}xecuting \textbf{A}s you \textbf{GE}ne\textbf{R}ate), a code execution framework to support script-style code generation.
We demonstrate this paradigm for Python, the dominant language in LLM code generation, though the principle applies to other interpreted languages.
Its core design follows a producer--consumer pipeline, where the LLM acts as the producer and the executor as the consumer.
As the language model generates tokens, an AST-based chunker incrementally identifies complete Python statements from the token stream and enqueues each chunk into a buffer.
Concurrently, an executor dequeues and runs available chunks within a persistent interpreter session, preserving variable bindings across successive batches.
The executor employs a dynamic batching strategy: when multiple chunks have accumulated in the queue, it merges them into a single batch, thereby amortizing the per-invocation setup overhead.
Additionally, \approach{} features an optional early error interruption mechanism: since chunks are executed as they are generated, a runtime error in any chunk is detected immediately, at which point \approach{} terminates the LLM generation and returns the error along with the partially generated code.
This avoids wasting generation time on code that depends on already-failed state.

We evaluate \textsc{Eager} across four Python benchmarks, seven LLMs, and three
execution environments.
The overlap
mechanism hides most execution behind generation.
It drives the non-overlapped
portion of execution time to near zero in most settings and lowers end-to-end
latency by up to 37.3\% on error-free runs, with the largest gains where
generation is fast relative to execution, so that execution makes up a larger share of total latency.
The optional early interruption avoids wasting generation on
already-failed code, and by handing the model earlier and more focused error
feedback it raises subsequent repair success by up to 44.3 percentage points on
data-centric benchmarks.
This prevents the model from being anchored by the incorrect code generated after the error, giving it more freedom to produce a corrected solution.

We open-source our artifacts at~\url{https://doi.org/10.6084/m9.figshare.31869469}. In summary, this paper makes the following contributions:
\begin{itemize}
\item We formalize parallel execution for LLM code interpreters and derive
  closed-form latency bounds that identify its speedup potential and the regimes where it pays off, including the condition under which it never regresses below serial execution, namely when the cumulative per-chunk overhead stays within the one-time serial setup cost.
  \item We present \textsc{Eager}, a Python implementation featuring AST-based
  chunking, gated dynamic batching, and early error interruption.
  \item We evaluate \textsc{Eager} on four benchmarks, seven LLMs, and three
  environments, reporting the latency benefit of overlap and the repair benefit
  of early interruption separately, along with the conditions under which each
  applies.
\end{itemize}

\section{Parallel Execution}
In this section, we introduce the general workflow of parallel execution and theoretically analyze its latency.

\subsection{Workflow}
\label{subsec:workflow}
Current LLMs follow a strictly sequential paradigm for code execution: the model first generates the entire program and only then invokes the execution environment.
This results in a clear temporal separation between generation and execution, leading to significant idle time on both sides.

We instead formulate the process as a \emph{streaming pipeline}.
The LLM acts as a \emph{producer} that autoregressively emits tokens.
A detection module continuously processes the token stream to identify executable chunks, defined as minimal syntactically complete and semantically executable units.
Once a chunk is detected, it is immediately dispatched to an execution engine, which maintains a persistent session to preserve program state.
This design enables temporal overlap across all stages: while the LLM generates tokens for later chunks, earlier chunks can already be detected and executed.

\subsection{Theoretical Modeling}
\label{sec:theory}
 
We model the system as a three-stage pipeline consisting of generation, detection, and execution. The total latency is determined by the critical path through these stages.
 
\noindent \textbf{Notation.}
We define:
\begin{itemize}
\item $L$: total number of tokens,
\item $v_{gen}$: generation speed (tokens per second),
\item $T_{FT}$: time-to-first-token,
\item $N$: number of executable chunks,
\item $l_i$: length of chunk $i$, with $\sum_{i=1}^N l_i = L$,
\item $\delta_i$: residual detection delay for chunk $i$ (the portion of detection cost not hidden behind generation),
\item $T_{setup}$: per-chunk execution overhead,
\item $T_{exe,i}$: execution time of chunk $i$.
\end{itemize}

\subsubsection{Serial Execution}
 
In the serial paradigm, the model first generates the complete program, after which the interpreter executes it as a monolithic block. 
The total latency is therefore:
\begin{equation}
T_{serial}
=
T_{FT} + \frac{L}{v_{gen}}
+ T_{setup}^{(\mathrm{full})}
+ T_{exe}^{(\mathrm{full})},
\end{equation}
where $T_{setup}^{(\mathrm{full})}$ denotes the one-time execution setup cost for the complete program, and $T_{exe}^{(\mathrm{full})}$ denotes the execution time of the full program.
 
For consistency with the chunked formulation, one may approximate:
\begin{equation}
T_{exe}^{(\mathrm{full})} \approx \sum_{i=1}^{N} T_{exe,i},
\end{equation}
but importantly, the serial baseline does not incur repeated per-chunk setup overhead or streaming detection overhead.

\subsubsection{Parallel Execution}
In the parallel paradigm, generation, detection, and execution are overlapped. Let $t_{g,i}$ denote the time when chunk $i$ has been fully generated, and let $t_{e,i}$ denote the time when execution of chunk $i$ completes.
 
\textbf{Generation.}
The time at which chunk $i$ has been fully generated depends on the cumulative token length of all preceding chunks:
\begin{equation}
t_{g,i} = T_{FT} + \frac{\sum_{j=1}^{i} l_j}{v_{gen}}
\end{equation}
 
\textbf{Detection.}
The detector processes the token stream online. We write the chunk-ready time as:
\begin{equation}
t_{d,i} = t_{g,i} + \delta_i
\end{equation}
where $\delta_i$ is the residual detection delay, i.e., the portion of detection cost that is not hidden behind generation.
 
\textbf{Execution.}
Chunk $i$ can only start executing once it has been detected from the generated code and the previous chunk has finished executing:
\begin{equation}
t_{e,i}
=
\max(t_{d,i},\, t_{e,i-1})
+ T_{setup} + T_{exe,i}
\end{equation}
 
\noindent The overall parallel latency is the completion time of the final chunk:
\begin{equation}
T_{parallel} = t_{e,N}
\end{equation}
 
\textbf{Closed-form characterization.}
Unrolling the recurrence gives
\begin{equation}
T_{parallel}
=
\max_{1 \le i \le N}
\left[
t_{g,i} + \delta_i + \sum_{j=i}^{N} (T_{setup} + T_{exe,j})
\right]
\end{equation}
 
\subsubsection{Latency Bounds}
 
The closed-form expression of the parallel execution allows us to derive upper and lower bounds on the latency.
 
\textbf{Upper bound.}
For any $i\in \{1, \ldots, N\}$, the generation prefix satisfies
$\sum_{j=1}^{i} l_j \le L$, the detection residual satisfies
$\delta_i \le \bar{\delta} \triangleq \max_{1 \le k \le N} \delta_k$, and the
execution tail satisfies
$\sum_{j=i}^{N}(T_{setup}+T_{exe,j}) \le \sum_{j=1}^{N}(T_{setup}+T_{exe,j})$.
Applying these to every term inside the outer maximum yields
 
\begin{equation}\label{eq:upper}
  T_{parallel}
  \;\le\;
  T_{FT} + \frac{L}{v_{gen}} + \bar{\delta}
  + N\,T_{setup} + \sum_{j=1}^{N} T_{exe,j}
\end{equation}
 
This upper bound corresponds to a \emph{zero-overlap} execution in which every
stage waits for its predecessor to complete entirely.  Compared with the serial
baseline, the additional cost is at most
\begin{equation}\label{eq:overhead}
  T_{parallel} - T_{serial}
  \;\le\;
  \bar{\delta} + N\,T_{setup} - T_{setup}^{(\mathrm{full})}
\end{equation}
, which captures the overhead from (i)~streaming detection and
(ii)~repeated per-chunk setup.  In practice $\bar{\delta}$ is on the order of
milliseconds because the detector operates on a lightweight grammar, so
the overhead is dominated by the cumulative setup cost
$N\,T_{setup} - T_{setup}^{(\mathrm{full})}$.
When both $\bar{\delta}$ and $N\,T_{setup} - T_{setup}^{(\mathrm{full})}$ are
negligible, the upper bound reduces to $T_{serial}$,  showing that the parallel scheme introduces no regression under these conditions.
 
\textbf{Lower bound.}
Two structural constraints yield complementary lower bounds.
 
\begin{enumerate}
\item \textbf{Generation constraint} (setting $i = N$).
The system must generate all tokens before the last chunk can complete:
\begin{equation}\label{eq:lb-gen}
  T_{parallel}
  \;\ge\;
  T_{FT} + \frac{L}{v_{gen}} + \delta_N + T_{setup} + T_{exe,N}
\end{equation}
 
\item \textbf{Execution constraint} (setting $i = 1$).
The system must execute all chunks in order after the first chunk becomes available:
\begin{equation}\label{eq:lb-exec}
  T_{parallel}
  \;\ge\;
  T_{FT} + \frac{l_1}{v_{gen}} + \delta_1
  + N\,T_{setup} + \sum_{j=1}^{N} T_{exe,j}
\end{equation}
\end{enumerate}
 
Combining both gives the composite lower bound:
\begin{equation}\label{eq:lb}
  T_{parallel}
  \;\ge\;
  \max\!\left\{
  \begin{aligned}
    & T_{FT} + \frac{L}{v_{gen}} + \delta_N + T_{setup} + T_{exe,N},\\
    & T_{FT} + \frac{l_1}{v_{gen}} + \delta_1 + N\,T_{setup}
      + \textstyle\sum_{j=1}^{N} T_{exe,j}
  \end{aligned}
  \right\}
\end{equation}
The first term dominates when generation is the bottleneck (\emph{generation-dominated regime}); the second dominates when execution is the bottleneck (\emph{execution-dominated regime}).
 
\subsubsection{Speedup Bounds}
 
Define the speedup $S = T_{serial} / T_{parallel}$.
From the lower bound~\eqref{eq:lb-gen} on $T_{parallel}$, the speedup is at most:
\begin{equation}\label{eq:speedup-ub}
  S
  \;\le\;
  \frac{T_{FT} + \frac{L}{v_{gen}} + T_{setup}^{(\mathrm{full})}
        + T_{exe}^{(\mathrm{full})}}
       {T_{FT} + \frac{L}{v_{gen}} + \delta_N + T_{setup} + T_{exe,N}}
\end{equation}
when $\delta_N$ and $T_{exe,N}$ are small relative to $L/v_{gen}$, which simplifies to
$S \lesssim 1 + (T_{setup}^{(\mathrm{full})} + T_{exe}^{(\mathrm{full})}) / (T_{FT} + L/v_{gen})$.
This is achieved when all execution is hidden behind generation.
From the upper bound~\eqref{eq:upper}, $S \ge 1$ whenever
$T_{setup}^{(\mathrm{full})} \ge \bar{\delta} + N\,T_{setup}$,
i.e., the one-time serial setup cost exceeds the cumulative
chunk overhead.
Whether this condition holds depends on the execution environment.
In our experimental setup, where chunks are dispatched to a persistent REPL session with per-call overhead on the order of 1\,ms, $T_{setup}$ and $\bar{\delta}$ are both small enough for the condition to be satisfied comfortably.
However, in environments with heavier per-chunk orchestration costs
(e.g., container cold starts or cross-process IPC), the cumulative
term $N\,T_{setup}$ may become non-negligible, and the condition
should be verified empirically.
 
\subsubsection{Regime Analysis under Uniform Chunks}
 
To obtain sharper insight, we analyze the special case of
\emph{uniform chunks}: $l_i = L/N$, $T_{exe,i} = \tau_e$, and
$\delta_i = \delta$.
Define the per-chunk generation time $\alpha \triangleq L/(N\,v_{gen})$
and per-chunk execution time $\beta \triangleq T_{setup} + \tau_e$.
The inner term of the closed-form expression becomes affine in $i$:
\begin{equation}
  f(i) = \underbrace{T_{FT} + \delta + (N+1)\beta}_{\text{constant}}
         + i\,(\alpha - \beta)
\end{equation}
yielding the following three regimes:
 
\textbf{R1: Generation-dominated ($\alpha > \beta$).}
The maximum is at $i = N$:
\begin{equation}\label{eq:gen-dom}
  T_{parallel} = T_{FT} + \frac{L}{v_{gen}} + \delta + \beta
\end{equation}
Execution of every chunk except the last is entirely hidden behind
generation.
 
\textbf{R2: Execution-dominated ($\alpha < \beta$).}
The maximum is at $i = 1$:
\begin{equation}\label{eq:exec-dom}
  T_{parallel} = T_{FT} + \alpha + \delta + N\beta
\end{equation}
Generation of every chunk except the first is hidden behind execution.
 
\textbf{R3: Balanced ($\alpha = \beta$).}
The pipeline is perfectly paced.
Setting $\alpha = \beta$, i.e., $L/(N\,v_{gen}) = T_{setup} + T_{exe}^{(\mathrm{full})}/N$, and solving for $N$ gives the critical chunk count:
\begin{equation}\label{eq:nstar}
  N^* = \frac{L/v_{gen} - T_{exe}^{(\mathrm{full})}}{T_{setup}}
\end{equation}
which is positive whenever total generation time exceeds total execution
time, the common case for LLM code generation.
For $N \le N^*$ additional chunks improve overlap;
beyond $N^*$ the cumulative setup overhead $N\,T_{setup}$ dominates
and latency degrades.

\begin{figure}[t]
\centering
\includegraphics[width=0.85 \linewidth]{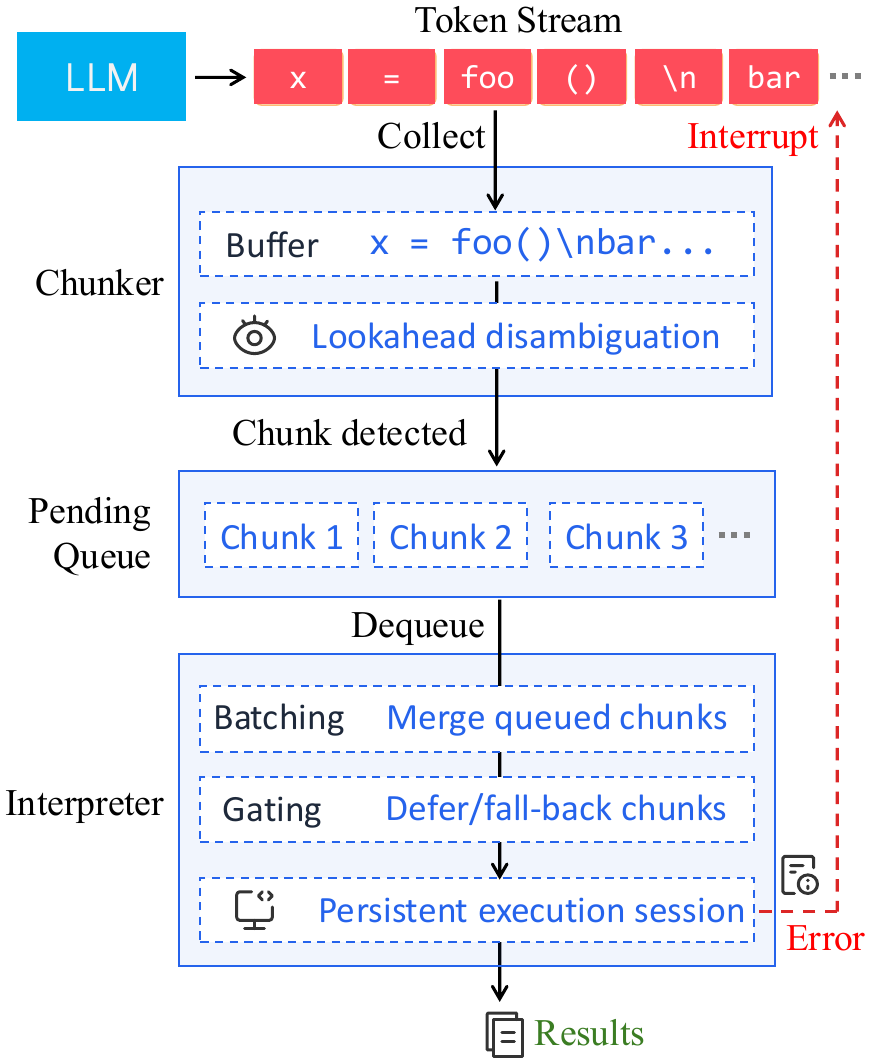}
\caption{Architecture of \approach{}.}
\label{fig:arch}
\end{figure}

\section{Implementation}
\label{sec:method}
Building on the theoretical framework in Section~\ref{sec:theory}, we present \approach{}, a concrete implementation of parallel execution for LLM code generation.
\approach{} instantiates the pipeline with design choices aimed at minimizing the detection and per-chunk setup overhead identified in Section~\ref{sec:theory}, without affecting execution outcomes.
 
As illustrated in Figure~\ref{fig:arch}, \approach{} consists of a chunker and an executor.
\emph{Chunker} accumulates the streaming tokens generated by LLMs in a buffer and identifies complete Python statements via AST parsing.
Detected chunks are dispatched to a \emph{pending queue}, from which an \emph{executor} dequeues, applies gating and batching optimizations, and delegates execution to a persistent session.
If a runtime error is encountered, the executor sends an interrupt signal back to terminate the LLM generation immediately (the dashed path in Figure~\ref{fig:arch}).
We describe each component in detail below.
 
\subsection{Producer: AST-Based Chunker}
 
The chunker operates on the streaming token output from an LLM and identifies executable chunks incrementally.
We implement the chunker based on AST statement boundaries.
As each token arrives, it is appended to a code buffer, and the chunker attempts to parse the buffer into a Python AST.
A chunk boundary is recognized when the buffer forms a complete top-level statement (e.g., assignments, loops, or function calls).
A statement is considered \emph{complete} when it can be unambiguously determined that no further tokens will be generated as part of it, i.e., the statement has no remaining portions yet to be produced by the LLM.
Once a chunk is confirmed as complete, the corresponding statements are removed from the buffer and dispatched to the executor, while any remaining tokens stay in the buffer for subsequent detection.
 
In many cases, syntactic completeness directly implies completeness.
For example, upon receiving \texttt{print("hi")} followed by a newline, the chunker can immediately confirm this as a standalone statement and dispatch it.
However, in other cases, a syntactically valid statement may still have remaining portions to be generated.
Consider generating a function where tokens arrive incrementally.
After receiving the first line of the function body, the parser can already yield syntactically valid results.
Yet the next line could continue the function body with additional statements, in which case the definition is not yet complete.
To handle such ambiguities, \approach{} employs a \emph{lookahead} strategy: when a statement parses successfully but its completeness cannot be definitively confirmed from syntax alone, the chunker waits for one additional token before committing.
If the next token rules out the possibility of the current statement continuing (e.g., by beginning a new top-level statement or indicating a dedent), the chunk is finalized and dispatched; otherwise, the buffer continues to accumulate.
This mechanism ensures that dispatched chunks are both syntactically complete and semantically independent.
 
\subsection{Consumer: Gated Executor with Dynamic Batching}\label{sec:gating}

The executor acts as the consumer of the pipeline, receiving chunks dispatched by the chunker.
It interacts with a persistent execution session, whether a local Python subprocess, a sandboxed environment, or a Docker container, that preserves all imports, variable bindings, and function definitions across successive chunk executions.
This execution session is therefore responsible for the execution of the chunks.

Confirmed chunks are placed into a pending queue.
When the executor becomes available, it merges all currently pending chunks into a single big chunk, rather than executing them one by one.
This \emph{dynamic batching} naturally adapts to the pace difference between the chunker and the executor: when execution is slower than detection, more chunks accumulate and are batched together, reducing the effective number of executor invocations and thus the cumulative setup cost $N\,T_{setup}$ in Equation~\ref{eq:upper}.
This also makes the speedup condition $S \ge 1$ easier to satisfy, since the overhead term $N\,T_{setup} - T_{setup}^{(\mathrm{full})}$ in Equation~\ref{eq:overhead} shrinks accordingly.

On top of batching, the executor applies a gating policy that classifies each chunk before dispatch.
The policy distinguishes four categories.
The first identifies chunks that are unprofitable to execute eagerly and defers them within the pipeline. The remaining three identify chunks whose \approach{} execution would be unsafe or would alter program semantics, and fall back to serial execution for them.
\begin{itemize}
\item \textbf{Low-yield chunks.}
Chunks whose execution time is negligible compared to the per-invocation setup overhead yield no useful overlap.
Typical examples are function or class declarations, which produce no observable computation when executed in isolation.
For such chunks, the setup cost $T_{setup}$ dominates the actual execution time $T_{exe,i} \approx 0$, so dispatching them individually only adds overhead.
The executor therefore defers these chunks: they remain in the pending queue and are merged with the next non-deferred chunk, at which point the declarations become available to support subsequent code that depends on them.

\item \textbf{External-side-effect chunks.}
\approach{} execution changes \emph{when} a statement runs relative to generation, not \emph{whether} it runs.
For most statements this reordering is harmless, but operations with irreversible, externally observable effects must not be committed before the surrounding code has been generated.
For chunks invoking process creation, filesystem mutation, network communication, or inter-process operations (e.g., \texttt{os}, \texttt{subprocess}, \texttt{shutil}, \texttt{socket}, \texttt{multiprocessing}), \approach{} disables overlap and executes the program serially, so that their side-effect timing matches the serial baseline.

\item \textbf{Timing-sensitive chunks.}
Some statements have semantics that depend on wall-clock timing or on the surrounding execution schedule, such as sleeping, timers, clock reads, signal delivery, and thread creation (e.g., \texttt{time}, \texttt{signal}, \texttt{threading}).
Executing such chunks ahead of generation would perturb a program's own timing or concurrency assumptions, altering its observable behavior.
\approach{} therefore falls back to serial execution for these programs, keeping such statements in program order and preserving correctness for code that observes its own timing.

\item \textbf{Dynamic-execution chunks.}
Dynamic execution entry points (\texttt{eval}, \texttt{exec}, \texttt{compile}, \texttt{\_\_import\_\_}) make a chunk's behavior undeterminable from its surface syntax, so the gating policy cannot reason about which of the categories above it belongs to.
Rather than dispatch code whose effects cannot be statically inspected, \approach{} conservatively falls back to serial execution.
\end{itemize}

The gate is applied as a static, name-based check over each chunk's parsed AST. Its default instance is a conservative denylist of standard-library modules that commonly carry external effects, and a deployer can extend this set for a given workload.

\subsection{Error Handling in the Pipeline}
 
The preceding subsections describe the normal flow of the pipeline: the chunker detects and dispatches chunks, and the executor batches and executes them.
We now describe how the pipeline handles runtime errors.
 
In serial execution, runtime errors are only observed after the entire program has been generated and executed as a whole.
In \approach{}, since chunks are executed incrementally alongside generation, a runtime error is caught as soon as the failing chunk finishes execution.
At that point, \approach{} immediately terminates the LLM generation process.
The error message, together with the code generated up to the point of failure, is returned to the caller.
No further tokens are generated or executed.
 
This early interruption provides two benefits.
First, it avoids spending generation time on code that depends on already-failed state, effectively reducing both the token count $L$ and chunk count $N$ on failure paths and thus lowering end-to-end latency.
Second, it delivers error feedback at the earliest possible point, enabling a tighter repair loop---as we empirically validate in RQ3 (Section~\ref{sec:rq3}), this earlier feedback leads to higher repair success rates in most scenarios.
 
Notably, this early interruption is an optional feature.
\approach{} can also be configured to continue generation after an error is detected and defer the error report until the full program has been produced.
Under this configuration, \approach{} still provides latency savings through overlapped execution of the error-free prefix, while preserving the same complete-code error semantics as serial execution.

\section{Experiment Setup}
\label{sec:eval}
Using \approach{} as a demonstration, we experimentally assess the performance of parallel execution.
In this section, we will introduce the settings of the experiments, including the benchmarks, LLMs, execution environments, and implementation details.
The rationale behind our setup is driven by answering the following three research questions:

\begin{itemize}
    \item \textbf{RQ1:} How much latency can \approach{} save compared to serial execution across different token generation speeds and execution environments?
    \item \textbf{RQ2:} Do the latency savings generalize from simulated to real LLM code generation?
    \item \textbf{RQ3:} Does the earlier error feedback provided by \approach{} help or hinder LLMs' subsequent code repair?
\end{itemize}

\subsection{Benchmarks}
We use four benchmarks where the LLMs are required to produce executable Python scripts.
\begin{itemize}
    \item \textbf{DSBench}~\cite{dsbench2024} and \textbf{DABench}~\cite{dabench2024} are both benchmarks that evaluate LLMs on data analysis tasks, where the model is given tabular data files along with natural language questions and must generate executable Python scripts to derive the answers. DSBench is sourced from Modeloff financial analysis competitions and Kaggle challenges, featuring 442 data analysis tasks with realistic settings such as long task descriptions, multimodal backgrounds, and multi-table structures. DABench provides 257 data analysis questions derived from real-world CSV files crawled from GitHub.

    \item \textbf{PandasPlotBench}~\cite{pandasplotbench2024} (PdPlotBench) is a human-curated benchmark for evaluating LLMs' ability to generate data visualization code. It contains 175 tasks, each requiring the model to produce plotting code given a natural language description and a Pandas DataFrame specification. 

    \item \textbf{GitChameleon}~\cite{gitchameleon2025} is a manually curated benchmark that evaluates LLMs' ability to generate version-specific Python code.
    It contains 116 code completion problems, each conditioned on a particular library version and accompanied by executable unit tests.
    Unlike conventional code generation benchmarks, GitChameleon specifically tests whether models can produce code compatible with a specified version of a library, reflecting real-world scenarios where developers are constrained to specific dependency versions.
\end{itemize}

\subsection{Large Language Models}
We evaluate \approach{} across seven LLMs spanning both open-weight and proprietary models, covering a range of model scales, speeds, and architectures.

For open-weight models, we use
DeepSeek-V3.2~\cite{deepseekv32}, a 685B-parameter MoE model with sparse attention;
MiMo-V2-Flash~\cite{mimov2flash}, Xiaomi's 309B MoE model (15B active) optimized for fast inference via hybrid sliding-window attention and multi-token prediction;
Qwen3-Coder~\cite{qwen3coder}, Alibaba's 480B MoE coding model (35B active) trained with long-horizon reinforcement learning for agentic coding;
and DeepSeek-Reasoner~\cite{deepseekr1}, a reasoning-focused model that employs extended chain-of-thought during inference.

For proprietary models, we use
GPT-4o-mini~\cite{gpt4omini}, OpenAI's lightweight multimodal model;
GPT-5.1-Codex-Mini~\cite{gpt51codexmini}, a variant of GPT-5.1-Codex further optimized for agentic coding tasks;
and Gemini-3.1-Flash-Lite~\cite{gemini31flashlite}, Google's model in the Gemini~3 series designed for high-volume, low-latency workloads.

\subsection{Execution Environments}
As the LLM-generated code could be executed in various environments
in practice, we experiment with three mainstream options:

\begin{itemize}
    \item \textbf{Local execution} runs the generated Python scripts directly on the host machine, offering the lowest overhead and fastest startup time. This represents the simplest setup commonly used in lightweight scripting and prototyping workflows.
    
    \item \textbf{Docker-based execution} runs the scripts inside an isolated Docker container~\cite{docker2024}. This provides a reproducible and sandboxed environment that prevents unintended side effects on the host system, and is widely adopted in production-grade agent frameworks for dependency management and security.

    \item \textbf{Open Interpreter sandbox} executes code through the Open Interpreter~\cite{openinterpreter2026} framework, which provides a managed sandboxed environment with built-in support for LLM-driven code execution. It represents a higher-level abstraction where the execution runtime is integrated into an end-to-end agent pipeline.
\end{itemize}
These three environments allow us to evaluate whether the latency
savings of \approach{} generalize across practical deployment settings.
All environments are configured with 2 CPU cores to reflect
the resource-constrained settings typical of cloud-hosted
code execution sandboxes.

\subsection{Evaluation Metrics}
We use two metrics to quantify the latency impact:
\begin{itemize}
    \item \textbf{Non-overlapped Execution Latency (NEL)} measures the portion of code execution time that falls outside the LLM generation phase.
In serial execution, the entire execution time is non-overlapped, as execution begins only after generation completes.
In parallel execution with \approach{}, part of the execution overlaps with the ongoing token generation, and this metric captures only the remaining portion that still contributes to user-perceived delay.
A lower non-overlapped execution time indicates that more execution has been effectively hidden behind generation.

\item \textbf{End-to-End Latency (E2EL)} measures the wall-clock time from the start of the LLM call to the completion of code execution, i.e., the total time the user must wait.
It equals the sum of the LLM generation time and the non-overlapped execution time.
This metric directly reflects the delay perceived by the user.

\end{itemize}


\subsection{Implementation Details}

All experiments are conducted on a server running Ubuntu 22.04 with an Intel Xeon Platinum 8352V processor.
To ensure reliable and reproducible timing measurements, we apply single-CPU isolation via \texttt{taskset} for local and Open Interpreter runs, and CPU pinning (\texttt{-{}-cpuset-cpus}) for Docker-based runs.
For code execution, the local and Docker environments use a persistent Python REPL subprocess that preserves imports, variables, and definitions across chunks, with per-call overhead of approximately 1\,ms.
The Open Interpreter environment uses its native Jupyter-kernel backend as the executor.
All experiments use isolated, task-local executors to prevent cross-task interference.
The LLMs are accessed through OpenRouter APIs~\cite{openrouter2025} via streaming mode.
All model names used throughout the paper correspond to their official API identifiers as of March 2026.

\begin{table}[t]
\centering
\setlength{\tabcolsep}{4pt}
\caption{Mock-token latency savings (\%) of \approach{} over serial execution across four benchmarks and three execution environments. NEL (Non-overlapped Execution Latency) saving measures the reduction in execution time that does not overlap with generation; E2EL (End-to-End Latency) saving measures the reduction in total wall-clock time.}
\begin{tabular}{@{}ll rr rr rr rr@{}}
\toprule
& & \multicolumn{2}{c}{DABench} & \multicolumn{2}{c}{DSBench} & \multicolumn{2}{c}{PdPlotBench} & \multicolumn{2}{c}{GitCham.} \\
\cmidrule(lr){3-4}\cmidrule(lr){5-6}\cmidrule(lr){7-8}\cmidrule(lr){9-10}
Env. & TPS & NEL & E2EL & NEL & E2EL & NEL & E2EL & NEL & E2EL \\
\midrule
Docker & 20  & 94.3 & 2.1  & \refine{96.3} & 1.1  & 88.9 & 6.1  & \refine{91.1} & \refine{4.1}  \\
Docker & 50  & 94.3 & 5.1  & \refine{95.7} & \refine{2.6}  & 88.4 & 13.7 & \refine{90.3} & \refine{8.3}  \\
Docker & 100 & 94.4 & 9.6  & \refine{94.5} & 5.1  & 88.0 & 23.4 & \refine{87.2} & \refine{12.8} \\
Docker & 200 & 93.9 & 17.4 & \refine{91.3} & 9.3  & 83.4 & 34.9 & \refine{82.8} & \refine{20.5} \\
\midrule
Local  & 20  & 93.8 & 2.2  & \refine{95.7} & 1.1  & 88.9 & 6.3  & \refine{88.3} & \refine{5.8}  \\
Local  & 50  & 93.9 & 5.3  & \refine{95.1} & 2.8  & 88.2 & 14.2 & \refine{84.5} & 10.6 \\
Local  & 100 & 94.0 & 10.3 & \refine{94.1} & 5.4  & 87.8 & 24.5 & \refine{85.3} & \refine{17.4} \\
Local  & 200 & 91.6 & 18.0 & \refine{89.8} & 10.2 & 81.1 & \refine{35.2} & \refine{73.4} & \refine{24.9} \\
\midrule
OI     & 20  & 94.3 & 3.2  & \refine{95.5} & 1.9  & 85.3 & 6.5  & \refine{92.5} & \refine{12.0} \\
OI     & 50  & 94.4 & 7.8  & \refine{95.0} & 4.4  & 82.8 & 14.5 & \refine{90.8} & \refine{23.2} \\
OI     & 100 & 94.3 & 14.6 & \refine{94.3} & 8.3  & 81.6 & 24.3 & \refine{88.7} & \refine{35.7} \\
OI     & 200 & 90.6 & 23.9 & \refine{91.6} & 15.0 & 73.3 & 34.0 & \refine{84.6} & \refine{48.3} \\
\bottomrule
\end{tabular}
\vspace{-4mm}
\label{tab:rq1-all}
\end{table}

\section{Results}
\subsection{Preliminary: Chunk Reconstruction Fidelity}

Before measuring the latency benefits of \textsc{Eager}, we first validate that its chunking mechanism preserves program integrity.
We collect the code generated by all seven LLMs across the four benchmarks (DABench, DSBench, PandasPlotBench, and GitChameleon) and run each program through \textsc{Eager}'s chunking pipeline under the Docker environment.
For every program, we concatenate the emitted chunks and compare the result against the original generated code.
Across all programs and all seven models, the reassembled code is character-level identical to the original in every case.
This confirms that the AST-based chunker correctly identifies statement boundaries without dropping, duplicating, or reordering any tokens, regardless of the generating model.

Since \textsc{Eager} executes chunks in sequence within a single persistent session, carrying forward all imports, variable bindings, and function definitions across chunk boundaries without any concurrency or re-initialization, lossless reconstruction directly implies execution equivalence for deterministic programs, which constitute all the LLM-generated code on our benchmarks.

\begin{table*}[t]
\centering
\small
\setlength{\tabcolsep}{3pt}
\caption{Latency comparison (ms) between serial execution (Baseline) and \approach{} (EAGER) across seven LLMs on Docker. NEL (Non-overlapped Execution Latency) measures time spent on execution that does not overlap with generation; E2EL (End-to-End Latency) measures total wall-clock time. Results are grouped by whether the generated code produces a runtime error. For GitChameleon error cases, EAGER NEL is marked ``---'' because \approach{} terminates generation upon detecting the error, leaving no post-generation execution phase.}
\begin{tabular}{@{}ll rr rr rr rr rr rr rr rr@{}}
\toprule
& & \multicolumn{8}{c}{Error-free (ms)} & \multicolumn{8}{c}{Error-encountered (ms)} \\
\cmidrule(lr){3-10}\cmidrule(lr){11-18}
& & \multicolumn{2}{c}{DABench} & \multicolumn{2}{c}{DSBench} & \multicolumn{2}{c}{PdPlotBench} & \multicolumn{2}{c}{GitCham.}
  & \multicolumn{2}{c}{DABench} & \multicolumn{2}{c}{DSBench} & \multicolumn{2}{c}{PdPlotBench} & \multicolumn{2}{c}{GitCham.} \\
\cmidrule(lr){3-4}\cmidrule(lr){5-6}\cmidrule(lr){7-8}\cmidrule(lr){9-10}
\cmidrule(lr){11-12}\cmidrule(lr){13-14}\cmidrule(lr){15-16}\cmidrule(lr){17-18}
Model & Exec. & NEL & E2EL & NEL & E2EL & NEL & E2EL & NEL & E2EL
      & NEL & E2EL & NEL & E2EL & NEL & E2EL & NEL & E2EL \\
\midrule
\multirow{2}{*}{DeepSeek-V3.2}
 & Baseline & 590 & 8467  & 256 & 28069 & 912 & 11182 & \refine{376} & \refine{14726}  & 638 & 9799  & 669  & 30362 & 678 & 15297 & 290 & 10161 \\
 & EAGER    & 2   & 7879  & 16  & 27816 & 199 & 10470 & \refine{28}  & \refine{14378}  & 0   & 3957  & 362  & 16547 & 0   & 7257  & ---  & \refine{8444}  \\
\midrule
\multirow{2}{*}{GPT-4o-mini}
 & Baseline & 607 & 3999  & 286 & 4887  & 759 & 5558  & \refine{337} & \refine{3589}   & 660 & 4991  & 1775 & 8264  & 811 & 6646  & 362 & 5766  \\
 & EAGER    & 1   & 3393  & 78  & 4679  & 54  & 4853  & \refine{41}  & \refine{3293}   & 0   & 1829  & 1447 & 4159  & 33  & 2377  & ---  & \refine{4947}  \\
\midrule
\multirow{2}{*}{MiMo-V2-Flash}
 & Baseline & 581 & 4082  & 272 & 35090 & 768 & 5004  & \refine{419} & \refine{3389}   & 658 & 4457  & 1625 & 14859 & 513 & 5983  & 424 & 4801  \\
 & EAGER    & 12  & 3513  & \refine{32}  & \refine{34848} & \refine{98}  & \refine{4334}  & \refine{62}  & \refine{3031}   & 5   & 2565  & 1316 & 6492  & \refine{43}   & \refine{3310}  & ---  & \refine{3810}  \\
\midrule
\multirow{2}{*}{Qwen3-Coder}
 & Baseline & 628 & 2858  & 156 & 7458  & 774 & 5347  & \refine{416} & \refine{5104}   & 696 & 4317  & 675  & 13163 & 675 & 5169  & 686 & 5247  \\
 & EAGER    & 80  & 2310  & 3   & 7296  & 81  & 4642  & \refine{66}  & \refine{4753}   & 85  & 1938  & 342  & 3289  & 0   & 2225  & ---  & \refine{3940}  \\
\midrule
\multirow{2}{*}{GPT-5.1-Codex}
 & Baseline & 594 & 1737  & 62  & 479   & 791 & 2244  & \refine{408} & \refine{1908}   & 631 & 1786  & 1259 & 6890  & 737 & 2912  & 429 & 2076  \\
 & EAGER    & 88  & 1231  & 34  & 451   & 190 & 1643  & \refine{121}  & \refine{1620}   & 127 & 832   & 979  & 6384  & 28  & 1515  & ---  & 1107  \\
\midrule
\multirow{2}{*}{Gemini-3.1}
 & Baseline & 583 & 1325  & 135 & 1381  & 753 & 1440  & \refine{414} & \refine{1674}   & 624 & 1546  & 259  & 1670  & 719 & 1558  & 378 & 1743  \\
 & EAGER    & 167 & 909   & 9   & 1255  & 215 & 903   & \refine{102}  & \refine{1362}   & 163 & 758   & 18   & 783   & 103 & 786   & ---  & \refine{1036}  \\
\midrule
\multirow{2}{*}{DeepSeek-R}
 & Baseline & 589 & 6073  & 153 & 15324 & 763 & 8467  & \refine{455} & \refine{7690}   & 661 & 7661  & 997  & 23680 & 681 & 10326 & 467 & 9994  \\
 & EAGER    & 1   & 5476  & 17  & \refine{15175} & \refine{47}  & \refine{7746}  & \refine{46}  & \refine{7281}   & 0   & 2934  & 699  & 10488 & 1   & 4619  & ---  & \refine{5247}  \\
\bottomrule
\end{tabular}

\label{tab:rq2-all}
\end{table*}

\subsection{RQ1: Latency Savings Across Generation Speeds and Environments}
As the generation speed of real LLMs are hard to systematically control, we choose a simulated setting.
Instead of using real LLMs to produce the token stream for \approach{} to execute, we use the already generated code solutions from all four benchmarks and replay them as a mock token stream at a fixed Token-Per-Second (TPS) rate.
Specifically, we use the gold (reference) solutions from PandasPlotBench and GitChameleon and the runnable solutions generated by DeepSeek-V3.2 for DABench and DSBench.
We sweep across four representative TPS rates---20, 50, 100, and 200---covering the range from slower open-weight model deployments to fast proprietary APIs.
We run all these code scripts across the three execution environments (local, Docker, and Open Interpreter) and compare \approach{} against serial execution in terms of non-overlapped execution time and end-to-end latency.
As shown in Table~\ref{tab:rq1-all}, \approach{} achieves consistently high NEL savings across all benchmarks, environments, and TPS rates, averaging 89.8\%.
This indicates that the vast majority of execution time is successfully hidden behind generation regardless of the specific configuration.
For example, on DABench and DSBench, the NEL savings remain above 89\% even at 200 TPS across all three environments, meaning that only about 10\% of the original execution time is left exposed after overlapping.
GitChameleon reaches 91.1\% NEL savings at 20 TPS under Docker, indicating that execution is largely hidden behind generation at lower generation speeds.
The E2EL savings, while more modest, follow a clear trend: they increase as generation speed decreases.
At 20 TPS, end-to-end savings are relatively small (1--6\% across benchmarks) because generation time dominates and the absolute execution time being hidden is small relative to the total.
At 200 TPS, the savings grow substantially (up to 34.9\% on PandasPlotBench under Docker), as faster generation compresses the generation window and makes execution a larger fraction of the total latency.
This is consistent with the theoretical prediction that the E2EL benefit of parallel execution scales with the ratio of execution time to generation time.
Across the three environments, the savings are broadly consistent, confirming that \approach{} generalizes across deployment settings.
Docker and Local show similar results, with Docker having a slight edge due to more stable timing from CPU pinning.
Open Interpreter exhibits marginally lower savings, particularly on PandasPlotBench and GitChameleon at higher TPS rates (e.g., 73.3\% NEL on PandasPlotBench at 200 TPS, but\ 83.4\% under Docker), which we attribute to the additional overhead of the Jupyter-kernel execution backend.
\begin{tcolorbox}[size=title]
{\textbf{Answer to RQ1:}} \approach{} hides 89.8\% of execution time behind generation on average across all tested configurations. The NEL savings are robust to changes in generation speed and execution environment, while the E2EL savings increase as generation speed grows, reaching up to 34.9\% at 200 TPS under Docker.
\end{tcolorbox}
\subsection{RQ2: Generalization to Real LLM Code Generation}
To validate whether the savings observed in RQ1 generalize to real-world settings, we use each of the seven LLMs to generate code for the four benchmark tasks via streaming.
The generated code is executed with \approach{} in the Docker environment, and the resulting latency is compared against the serial execution baseline.
Notably, when a runtime error occurs during execution,
\approach{} interrupts the code generation process immediately.
We therefore report results separately for error-free and error-encountered executions.
As shown in Table~\ref{tab:rq2-all}, \approach{} consistently reduces latency across nearly all model--benchmark combinations under both conditions.
For error-free executions, the non-overlapped execution latency (NEL) drops to near zero in many cases (e.g., 2\,ms for DeepSeek-V3.2 on DABench, 1\,ms for GPT-4o-mini on DABench), indicating that execution is almost entirely hidden behind generation for these settings.
The end-to-end latency (E2EL) improves correspondingly, with reductions ranging from modest savings on benchmarks where execution is already fast (e.g., DSBench with GPT-5.1-Codex, where the baseline NEL is only 62\,ms) to substantial reductions where execution constitutes a larger fraction of total latency (e.g., Gemini-3.1 on PandasPlotBench, from 1440\,ms to 903\,ms, a 37.3\% reduction).
The NEL savings are largest for slower models, whose longer generation window leaves more room to hide execution.
The E2EL savings move the other way: they are largest for faster models such as Gemini-3.1, where the shorter generation window makes execution a bigger share of total latency. 
This matches the speedup bound in Section~\ref{sec:theory}, where the end-to-end benefit grows as generation time falls relative to execution.
For error-encountered executions, \approach{} provides even larger latency reductions, as the early interruption mechanism eliminates both the remaining generation time and the execution of subsequent code that would inevitably depend on already-failed state.
This is most visible in the E2EL:
Qwen3-Coder on DSBench drops from 13163\,ms to 3289\,ms,
and DeepSeek-R on PandasPlotBench achieves a 55.3\% reduction.
The NEL for error cases is often 0\,ms on benchmarks like DABench and PandasPlotBench, meaning the error was caught entirely within the generation window.
For GitChameleon error cases, the NEL under \approach{} is undefined because generation is terminated at the point of error, so only E2EL is reported.
The E2EL nonetheless shows clear improvements across all models (e.g., DeepSeek-R from 9994\,ms to 5247\,ms), confirming that early interruption saves substantial waiting time even when the post-generation tail cannot be measured.
\begin{table*}[t]
\centering
\setlength{\tabcolsep}{4pt}
\renewcommand{\arraystretch}{1.05}
\caption{Error resolution rates (\%) for repairing from partially generated code (Partial) versus complete code (Full) across four benchmarks. $\Delta$ denotes the percentage-point difference.}
\begin{tabular}{@{}l ccl ccl ccl ccl@{}}
\toprule
& \multicolumn{3}{c}{DABench} & \multicolumn{3}{c}{DSBench} & \multicolumn{3}{c}{PandasPlotBench} & \multicolumn{3}{c}{GitChameleon} \\
\cmidrule(lr){2-4}\cmidrule(lr){5-7}\cmidrule(lr){8-10}\cmidrule(lr){11-13}
Model & Full & Partial & $\Delta$ & Full & Partial & $\Delta$ & Full & Partial & $\Delta$ & Full & Partial & $\Delta$ \\
\midrule
DeepSeek-V3.2  & 48.4 & 64.5 & +16.1          & 71.3 & 74.4 & +3.0  & 61.5 & 69.2 & +7.7           & \refine{66.0} & \refine{66.0} & \refine{+0.0}  \\
GPT-4o-mini    & 67.1 & 74.7 & +7.6           & 32.3 & 76.6 & \textbf{+44.3} & 64.8 & 74.1 & +9.3  & \refine{33.8} & \refine{35.1} & +1.3  \\
MiMo-V2-Flash  & 57.9 & 71.1 & +13.2          & 51.8 & 62.8 & +10.9 & \refine{52.9} & \refine{70.6} & \refine{+17.6}          & \refine{45.7} & \refine{42.9} & \refine{-2.9}  \\
Qwen3-Coder    & 37.1 & 69.4 & \textbf{+32.3} & 60.3 & 78.7 & +18.4 & 58.3 & 83.3 & \textbf{+25.0} & \refine{48.9} & \refine{46.8} & \refine{-2.1}  \\
GPT-5.1-Codex  & 78.7 & 80.9 & +2.1           & 31.0 & 36.1 & +5.1  & 81.8 & 86.4 & +4.5           & 82.5 & 67.5 & -15.0 \\
Gemini-3.1     & 43.4 & 75.5 & \textbf{+32.1} & 70.0 & 89.1 & +19.1 & 78.4 & 86.5 & +8.1           & \refine{57.4} & \refine{57.4} & +0.0  \\
DeepSeek-R     & 59.0 & 77.0 & +18.0          & 53.5 & 68.6 & +15.1 & 73.7 & 94.7 & \textbf{+21.1} & \refine{68.6} & \refine{60.8} & \refine{-7.8}  \\
\bottomrule
\end{tabular}

\label{tab:rq3-repair}
\end{table*}

\begin{tcolorbox}[size=title]
{\textbf{Answer to RQ2:}} The latency savings observed under simulated conditions in RQ1 generalize to real LLM outputs across all seven models and four benchmarks. For error-free executions, \approach{} reduces the NEL to near zero in most settings, cutting E2EL by up to 37.3\%, while for error-encountered executions the early interruption mechanism delivers even larger savings of up to 75.0\%.
\end{tcolorbox}
\subsection{RQ3: Effect of Earlier Error Feedback on Code Repair}\label{sec:rq3}
As observed in RQ2, \approach{} catches errors during generation rather than after it completes, providing earlier feedback to the LLM.
This early interruption also terminates the code generation process, leaving the remaining code ungenerated.
In this RQ, we investigate whether the partial code resulting from early error interruption affects the model's ability to resolve the detected error in subsequent repair attempts.
Specifically, for each error-encountered sample in RQ2, we append the error message to the already-generated code and let the same model continue generating a fix.
We compare two conditions (\textbf{Full} and \textbf{Partial}): under serial execution, the model continues with the complete code followed by the error message; under \approach{}, the model continues with the partially generated code followed by the error message.
We then compare the error resolution rates between the two conditions—that is, whether the repaired code executes without reproducing the original runtime error.
As reported in Table~\ref{tab:rq3-repair}, across the three data-centric benchmarks (DABench, DSBench, and PandasPlotBench), repairing from partial code consistently achieves higher error resolution rates than repairing from full code, with improvements ranging from +2.1 to +44.3 percentage points.
The gains are particularly pronounced for models such as Qwen3-Coder and Gemini-3.1, which achieve over 30 percentage points of improvement on DABench and GPT-4o-mini, which gains +44.3 percentage points on DSBench.
This result may seem counterintuitive at first, as one might expect that having the complete code would provide the model with more context for repair.
However, we hypothesize that the full program, which has already executed to completion and failed, may anchor the model toward preserving its original (flawed) logic.
In contrast, the partial code from \approach{} leaves the remainder ungenerated, giving the model more freedom to regenerate a corrected solution from the point of failure.
Moreover, we observe that on the three data-centric benchmarks, errors tend to occur early in the generated code, at a median position of around 35\% of the total code length.
This means that under serial execution, on average 65\% of the code is generated \emph{after} the error has already occurred, without any awareness of the runtime failure.
This post-error suffix often depends on state that no longer holds
(e.g., referencing a DataFrame column that caused a \texttt{KeyError}), providing the model with misleading context during repair.
By contrast, \approach{} interrupts generation at the point of failure,
avoiding this misleading suffix entirely.
The exception is GitChameleon, where repairing from partial code slightly underperforms in most cases.
We attribute this to the nature of GitChameleon tasks: they are version-specific code completion problems where the error typically stems from using an incorrect API for the target library version.
In these tasks, the code after the error point often encodes the \emph{intended} API usage, which is precisely the information needed to diagnose a version mismatch.
To investigate further, we analyzed the 27 cases (across all models) where full-code repair resolved the error but partial-code repair did not: 63\% of these cases had their code truncated by early interruption (tokens saved more than 0\%), confirming that the lost suffix contained error-relevant context.
The remaining 37\% showed identical input code, where the difference is attributable to LLM repair stochasticity.
\begin{tcolorbox}[size=title]
{\textbf{Answer to RQ3:}} The early interruption behavior of \approach{} can be advantageous for subsequent code repair. On three data-centric benchmarks, repairing from partial code improves error resolution rates by 2.1 to 44.3 percentage points over repairing from the complete program. The exception is GitChameleon, where the truncated suffix contains version-specific API context needed for repair.
\end{tcolorbox}

\section{Threats to Validity}
\label{sec:threats}

\noindent \textbf{Language generalizability}.
Our implementation and evaluation target Python. This is a natural rather than arbitrary choice, since script-style generation in interpreted languages is the dominant setting for the code-interpreter and code-as-action systems that parallel execution targets (Section~\ref{sec:intro}).
The pipeline itself is language-agnostic, and only the execution backend depends on a REPL-style interpreter, so extending to other interpreted languages such as JavaScript or R is straightforward.
Compiled languages would require a different backend (e.g., incremental compilation or JIT-based evaluation), which raises execution-model questions, such as whether compilation can itself be overlapped with generation, that are orthogonal to the scheduling mechanism studied here.
We leave this to future work.
Regardless of how that backend is realized, Equation~\ref{eq:overhead} bounds the worst case: parallel execution does not regress below serial execution as long as the cumulative per-chunk overhead stays within the serial setup cost, the condition met in our REPL setup, where we observe no regression in any configuration.

\noindent \textbf{Benchmark representativeness}.
Our method targets single-file scripts, which are the intended application domain of EAGER and the dominant workload of LLM code interpreters in both industry and research.
This is a deliberate scope rather than an incidental restriction: parallel execution yields its largest benefit precisely in this generation-dominated regime (Section~\ref{sec:theory}), where generation time dominates lightweight per-statement execution.
Tasks dominated by heavy execution, such as repository-scale agent tasks that compile a project or run a full test suite, fall into the execution-dominated regime and lie outside EAGER's intended scope. By Equation~\ref{eq:overhead}, EAGER does not regress below serial execution in that regime as long as the same overhead condition holds, but its latency savings are small there, so we do not target such settings.

\noindent \textbf{Docker cold-start outliers}.
In the Docker environment, the tasks in GitChameleon require creating
isolated virtual environments for version-specific library testing.
A small number of tasks (46 out of the 990 tasks across all four benchmarks) exhibited abnormally high
execution times due to virtual environment cold-start overhead.
We exclude these outliers from our reported results.
Removing this filter changes the aggregate results by less than
1 percentage point, confirming that our conclusions are robust to this decision.
\section{Related Work}
\label{sec:related}

\noindent\textbf{Code Generation with Execution in the Loop}
Recent LLM work has used code execution as an important mechanism for reasoning and interaction.
For example, program-aided reasoning methods such as PAL~\cite{Gao22} and Program of Thoughts~\cite{Che22} usually let the model first generate a complete program and only then execute it, using code mainly as a reliable substrate for arithmetic or symbolic computation rather than as a source of fine-grained feedback during decoding . Related language-to-code approaches in the same vein also largely follow this generate-then-execute pattern~\cite{Shi22}. Follow-up work preserved this basic structure while making the outer loop more execution-aware. LEVER uses execution outcomes to verify and rerank candidate programs~\cite{ni2024lever}, while Chain of Code augments executable code with selective emulation when some steps cannot be directly executed~\cite{Li23}. In parallel, self-correction systems such as Self-Debugging~\cite{Che23}, CYCLE~\cite{Din24}, and runtime-verification-based debugging methods~\cite{Zho24} treat execution errors, failed tests, or traces as signals for iterative repair after an initial solution has already been produced.
More recently, EG-CFG~\cite{lavon2025execution} tests the code line-by-line as it is being written, using real-time execution feedback to steer the model toward functionally correct and executable solutions.
Our work is orthogonal to these efforts:
rather than improving code quality, we reduce user-perceived latency by overlapping generation and execution, a technique that composes naturally with them.

\noindent\textbf{Execution Environments for LLMs}
Recent LLM systems have also developed increasingly capable execution substrates, including interactive interpreters, notebook-style runtimes, and containerized sandboxes that make generated code executable, stateful, and reproducible across turns. OpenCodeInterpreter integrates code generation with execution and iterative refinement in a code-interpreter-style workflow~\cite{Zhe24}, while InterCode exposes execution as part of the task environment through self-contained Docker sandboxes and standardized feedback channels~\cite{yang2024intercode}. Concurrently, infrastructure-oriented work such as MPL-Sandbox emphasizes practical multi-language isolation and unified compiler or runtime feedback for LLM-based coding systems~\cite{Dou24b}, and notebook-centric agents and benchmarks further show the importance of maintaining persistent execution state in data analysis settings~\cite{Yin22,Hon24}. These systems make execution available to the model, but their main goal is to provide a reliable runtime for interaction, evaluation, or iterative repair, rather than to minimize end-to-end response time.
In contrast, our focus is not on building a richer sandbox or stronger debugging loop, but on a systems-level scheduling question: how to overlap generation and execution so that runtime feedback can be exploited without forcing the user to wait for a strictly sequential generate-then-run pipeline.

\section{Discussion}
\label{sec:discussion}


\noindent\textbf{Implications for programming language design.}
Existing programming languages were designed under the assumption that
source code is written in its entirety before execution.
However, LLM-based code generation fundamentally changes this assumption:
code is produced token by token as a stream, with each prefix
potentially forming a meaningful partial program.
This mismatch forces systems like \approach{} to reconstruct executability
from a stream that the language was never designed to support incrementally---relying
on AST parsing heuristics, lookahead strategies, and gating policies to
determine when a partial program is safe and worthy to execute.
As LLM-generated code becomes an increasingly prevalent mode of program creation,
we see an opportunity for future programming languages to treat
\emph{streamability} as a first-class design goal:
for instance, through explicit statement delimiters that eliminate boundary ambiguity,
or through language-level primitives that allow the runtime to consume and execute
code incrementally as it is produced.
Such designs would reduce the complexity of systems like \approach{}
and more broadly benefit the emerging ecosystem of AI-assisted programming.


\noindent\textbf{Preserving execution semantics under overlap}
\approach{} dispatches code before the full program is available, which raises a natural concern: running statements ahead of generation may change the program behavior that serial execution would not?
\approach{} mitigates this issue through its gating policy in Section~\ref{sec:gating}.
It routes statements whose observable behavior may depend on \emph{when} they run, such as those with sensitivity to wall-clock timing and scheduling, onto the serial path, where they are committed only after all program has been generated.
Overlap is therefore confined to computation that is insensitive to its position relative to generation, where reordering cannot produce an externally observable consequence absent under serial execution.
In this sense, \approach{} preserves the observable semantics of the serial baseline.

It is worth emphasizing that \approach{} is a scheduling layer deciding when a statement runs relative to generation.
It is not an isolation mechanism and does not govern what generated code is permitted to do.
Confining untrusted code remains the responsibility of the execution environment, whether a container, a restricted subprocess, or a managed sandbox, exactly as in the serial setting.
Because effect-bearing statements retain their serial timing, an environment that behaves correctly under serial execution behaves identically under \approach{}: it inherits whatever isolation the deployment already provides without weakening it.
The gating check itself is a static, conservative over-approximation, and any statement it cannot classify as position-insensitive runs serially.
We treat the completeness of that classification as separate from the latency mechanism studied here.

\section{Conclusion}
\label{sec:conclusion}

In this paper, we introduced parallel execution, a paradigm that dispatches LLM-generated code statements to the interpreter as they are produced, overlapping generation and execution to reduce end-to-end latency.
We formalized this paradigm theoretically and realized it in \approach{}, a framework featuring AST-based chunking, dynamic batching, and early error interruption.
Experiments across four benchmarks, seven LLMs, and three execution environments showed that \approach{} consistently hides the majority of execution time behind generation, achieving end-to-end latency reductions of up to 37.3\%, while the early error interruption mechanism improved error resolution rates by up to 44.3 percentage points

\bibliographystyle{IEEEtran}
\bibliography{main}

\end{document}